\documentclass[rnote]{aa} % for the research notes referee
\usepackage{natbib}
\bibpunct{(}{)}{;}{a}{}{,}
\usepackage{graphicx}
\usepackage{txfonts}
\begin{document}

\title{Determining Parameters of Cool Giant Stars by Modeling Spectrophotometric and Interferometric Observations Using the \textsc{SAtlas} Program}
\author{Hilding R. Neilson\inst{1} \and John B. Lester\inst{2}}
\institute{Department of Astronomy \& Astrophysics, University of Toronto, neilson@astro.utoronto.ca \and Department of Chemical and Physical Sciences, University of Toronto Mississauga, lester@astro.utoronto.ca}

\date{Received / Accepted}
\abstract{Optical interferometry is a powerful tool for observing the intensity structure and angular diameter of stars. When combined with spectroscopy and/or spectrophotometry, interferometry provides a powerful constraint for model stellar atmospheres.}
{The purpose of this work is to test the robustness of the spherically symmetric version of the \textsc{Atlas} stellar atmosphere program, \textsc{SAtlas}, using interferometric and spectrophotometric observations.}
{Cubes (three dimensional grids) of model stellar atmospheres, with dimensions of luminosity, mass, and radius,  are computed to fit observations for three evolved giant stars, $\psi$ Phoenicis, $\gamma$ Sagittae, and $\alpha$ Ceti. The best--fit parameters are compared with previous results.}
{The best--fit angular diameters and values of $\chi^2$ are consistent with predictions using \textsc{Phoenix} and plane--parallel \textsc{Atlas} models.   The predicted effective temperatures, using \textsc{SAtlas},  are about $100$ to $200$ $\rm{K}$ lower, and the predicted luminosities are also lower due to the differences in effective temperatures.}
{It is shown that the \textsc{SAtlas} program is a robust tool for computing models of extended stellar atmospheres that are consistent with observations.  The best--fit parameters are consistent with predictions using \textsc{Phoenix} models, and the fit to the interferometric data for $\psi$ Phe differs slightly, although both agree within the uncertainty of the interferometric observations.}


\titlerunning{Testing the \textsc{SAtlas}  Program with Interferometric Observations}
\authorrunning{Neilson \& Lester}
\maketitle

\section{Introduction}
Optical interferometry accurately measures the combination of the angular diameter and the structure of the intensity distribution of a stellar disk.  The combination of interferometry with spectroscopy and/or spectrophotometry is a powerful tool for determining effective temperatures and other properties of stars, and the growth of optical interferometry is testing theoretical models of stellar atmospheres in new ways.  

In a series of articles, \cite{Wittkowski2004, Wittkowski2006b, Wittkowski2006a} fit Very Large Telescope Interferometer (VLTI) observations of cool giants using the VINCI instrument with spherically symmetric stellar atmosphere models in local thermodynamic equilibrium (LTE) computed with the \textsc{Phoenix} program \citep{Hauschildt1999}, and with plane parallel models from \textsc{Phoenix} and \textsc{Atlas} \citep{Kurucz1970, Kurucz1993}.  The authors demonstrated that optical interferometry can detect the wavelength--dependent limb--darkening of cool giants and that the limb--darkening can be used to constrain model stellar atmospheres.  These works are based on observations of three stars: $\psi$ Phoenicis (M4 III), $\gamma$ Sagittae (M0 III), and $\alpha$ Ceti (M1.5 III) also called Menkar. 

The purpose of this note is to model these interferometric observations using the new spherically symmetric version of the \textsc{Atlas} program (\textsc{SAtlas}) developed by \cite{Lester2008} and to determine fundamental parameters of the three stars.  There are two versions of the \textsc{SAtlas} program, one using opacity distribution functions  while the other uses opacity sampling; in this work we use the first version.  The radiation field is computed using the method suggested by \cite{Rybicki1971}, which is a reorganization of the \cite{Feautrier1964} method.

In this work we compute cubes (three dimensional grids) of spherically symmetric stellar atmospheres with dimensions of luminosity, mass, and radius to model interferometric visibilities to fit the VLTI/VINCI observations and broadband spectrophotometry from \cite{Johnson1975}.  This will provide a robust test of the program and a comparison to the results predicted using spherically symmetric \textsc{Phoenix} models.  In the next section, we outline the method for computing synthetic visibilities and predicting radii, effective temperatures, luminosities, masses, and gravities.  We present the results in the third section and the discussion in the fourth section.
 
\section{Method for Computing Visibilities and Determining Stellar Parameters}
The goal of this note is to determine the global properties of the three observed stars.  Interferometric observations are ideal for measuring the angular diameter, and with the HIPPARCOS parallax \citep{Perryman1997} one can find the linear radii of the stars.  The visibility \citep{Davis2000, Tango2002}  at wavelength $\lambda$ is
\begin{equation}\label{e1}
V_{\rm{LD}}(\lambda) = \int_0^1S_\lambda I_\lambda(\mu)J_0\left[\pi \theta_{\rm{LD}}(B/\lambda)(1 - \mu^2)^{1/2}\right] \mu \rm{d}\mu.
\end{equation}
In this equation, $S_\lambda$ is the sensitivity function of the instrument (VINCI in this case) \citep{Kervella2000}, $J_0$ is the zeroth--order Bessel function, $B$ is the length of the baseline of the interferometer, and $\mu$ is the cosine of the angle between the center of the stellar disk and a distance from the center.  The center of the disk corresponds to $\mu = 1$ and the edge of the disk is $\mu = 0$ as seen from the center of the star.   To match the broadband VLTI/VINCI observations, the monochromatic model visibilities must be integrated and squared to get the squared visibility amplitudes,
\begin{equation}\label{e2}
|V_{\rm{LD}}|^2 = \frac{\int_0^\infty |V_{\rm{LD}}(\lambda)|^2 \rm{d}\lambda}{\int_0^\infty S_\lambda^2F_\lambda^2 \rm{d}\lambda}.
\end{equation}
The denominator of Eq. \ref{e2} normalizes the visibilities, where $F_\lambda$ is the flux of model.  For a given model atmosphere, we can find the best--fit angular diameter with a minimum $\chi^2$ value.

The next step is to use the model stellar atmospheres to fit spectrophotometric observations.  The purpose of fitting spectrophotometry is to constrain the effective temperature and the stellar flux; however, the effective temperature is dependent on the angular diameter.  The modeled flux is scaled using the same method and assumptions as in \cite{Wittkowski2004} to fit the broadband data from \cite{Johnson1975}. The best--fit effective temperature is determined using the best--fit angular diameter from the interferometric fit to break the degeneracy.
   \begin{figure}
   \centering
   \includegraphics[]{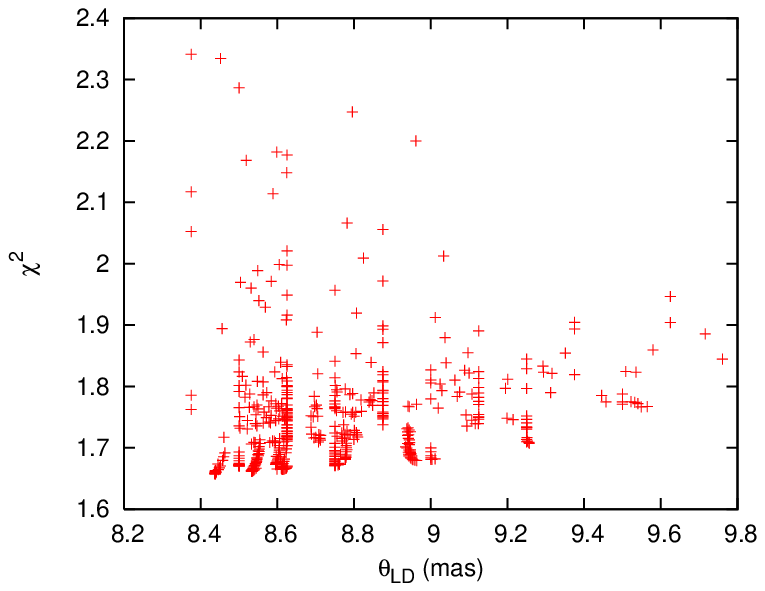}
    \includegraphics[]{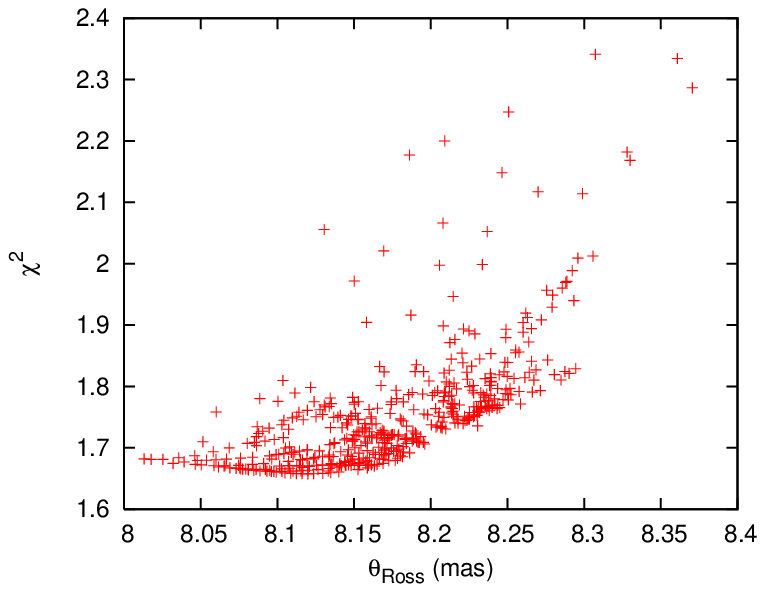}
      \caption{Values of minimum $\chi^2$ for each model stellar atmosphere for fitting the interferometric data of $\psi$ Phe as a function of (Top) $\theta_{\rm{LD}}$ and (Bottom) $\theta_{\rm{Ross}}$.  
              }
         \label{f1}
   \end{figure}
   
The linear radius of the star is determined using the best--fit angular diameter and HIPPARCOS parallax, and the luminosity is calculated using the radius and the effective temperature.  The mass and gravity are constrained by comparing the predicted effective temperature and luminosity with theoretical stellar evolution tracks \citep{Girardi2000}. 

\section{Results}
\begin{table}
\begin{center}
\caption{Best--Fit Parameters of the Three Stars}\label{t1}

\begin{tabular}{lccc}
\hline
& $\psi$ Phe & $\gamma$ Sge & $\alpha$ Cet \\
\hline
$\theta_{\rm{LD}} (\rm{mas})$ & $8.7\pm 0.3$ & $6.3 \pm 0.1$ & $12.6 \pm 0.2$ \\
$\theta_{\rm{Ross}} (\rm{mas})$ & $8.13 \pm 0.1$ & $6.02 \pm 0.1$ & $12.10 \pm 0.05$ \\
$\chi^2$ & $1.66$ & $0.634$ & $0.98$ \\
$\pi (\rm{mas})$ & $10.15 \pm 0.15$ & $11.9 \pm 0.71$ & $14.82 \pm 0.83$ \\
$R(R_\odot)$ & $85 \pm 1.6$& $54 \pm 4$ & $88 \pm 5$ \\
$T_{\rm{eff}} (\rm{ K})$ & $3415 \pm 87$ & $3650 \pm 78$ & $3584 \pm 166$ \\
$L(L_\odot)$ & $882 \pm 96$ & $468 \pm 81$ & $1147 \pm 249$\\
$M(M_\odot)$ & $0.85 \pm 0.1$ & $0.9 \pm 0.2$ & $1.3 \pm 0.4$ \\
$\log g$ (cm s$^{-2})$ & $0.51^{+0.06}_{-0.02}$ & $0.93^{+0.15}_{-0.17}$ &  $0.66^{+0.17}_{-0.20}$\\
\hline
\end{tabular}
\end{center}
\end{table}
For each of the three stars, we compute a cube of models with luminosity, mass, and radius as input parameters.  The metallicity is assumed to be solar, consistent with the conclusions of \cite{Feast1990}, and the microturbulence is zero.  These assumptions can be tested, but for this work variations of metallicity and microturbulence are ignored.   The range of values for the input mass, luminosity and radius are chosen based on the results of \cite{Wittkowski2004, Wittkowski2006b, Wittkowski2006a}.  

   \begin{figure}
   \centering
   \includegraphics[]{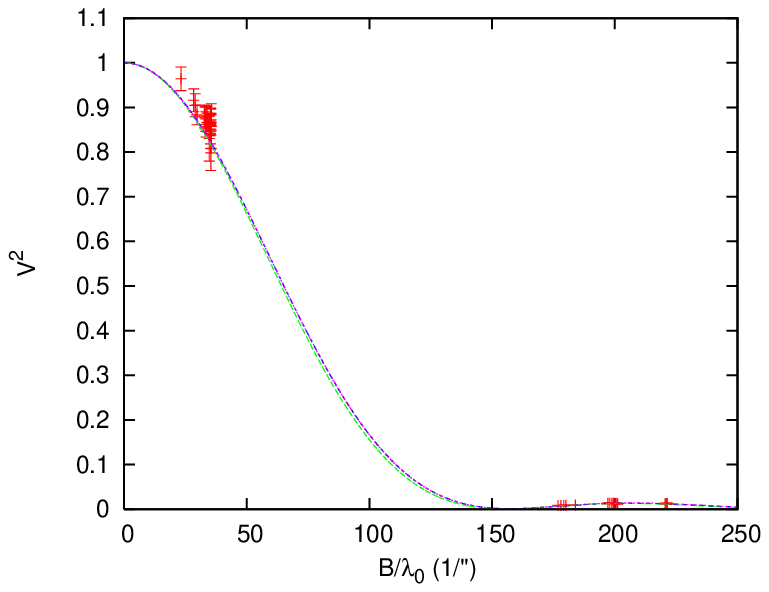}
    \includegraphics[]{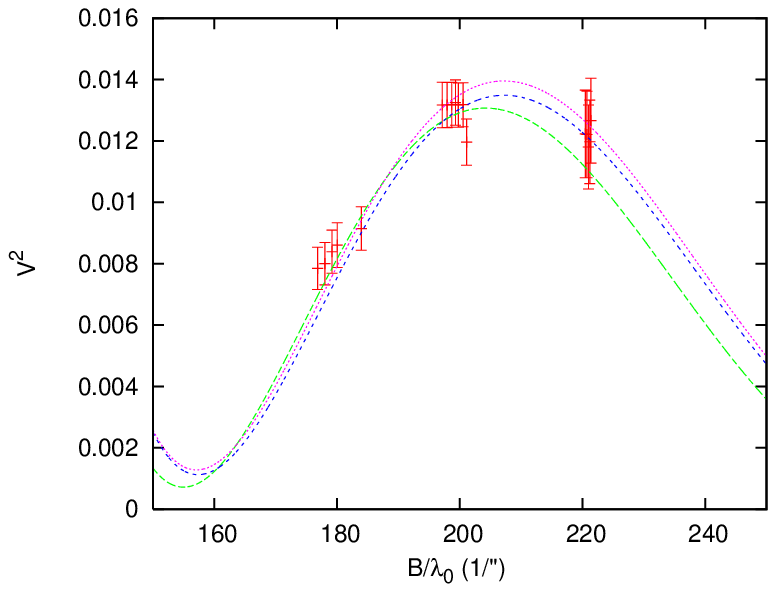}
      \caption{(Top) The visibilities calculated for three \textsc{SAtlas} models and the interferometric data for $\psi$ Phe .  (Bottom) Close--up of the second lobe  of the visibility amplitude.  The dotted line refers to the model stellar atmosphere for $L = 1580L_\odot$, M = $1.6M_\odot$, and $R = 60R_\odot$, the dot--dashed line represents a model with $L = 980L_\odot$, M = $1.2M_\odot$, and $R = 80R_\odot$, and the dashed line is for $L = 630L_\odot$, M = $0.6M_\odot$, and $R = 120R_\odot$. 
              }
         \label{f2}
   \end{figure}

For $\psi$ Phe, the mass range is chosen to be $0.6$ to $1.6M_\odot$ in steps of $0.2M_\odot$; the luminosity range is $630$ to $1580L_\odot$ in steps of $50L_\odot$ and the radius range is $60$ to $120R_\odot$ in steps of $20R_\odot$.  There are $480$ models computed for $\psi$ Phe.  The mass range for $\gamma$ Sge is $1.0$ to $1.9M_\odot$ in steps of $0.1M_\odot$, the luminosity range is $400$ to $700L_\odot$ in steps of $50L_\odot$ and the radius range is $30$ to $70R_\odot$ in steps of $10R_\odot$, leading to $350$ models for this star.  There are $1680$ models computed for $\alpha$ Cet to provide a larger range of parameters and a more robust test of the \textsc{SAtlas} program.  The mass range is $1.5$ to $3.0M_\odot$ in steps of $0.1M_\odot$, the luminosity range is $10^3$ to $2\times 10^3L_\odot$ in steps of $50L_\odot$ and the radius range is $60$ to $100R_\odot$ in steps of $10R_\odot$.

We fit the limb--darkened angular diameter, $\theta_{\rm{LD}}$, to the interferometric observations of each star and find a minimum value of $\chi^2$.  The limb--darkened angular diameter is the angular diameter to the edge of the stellar disk corresponding to the the layer where $\mu = 0$, and from the stellar atmosphere models we determine the Rossland angular diameter by multiplying $\theta_{\rm{LD}}$ by the ratio of radius of the stellar atmosphere model at  $\tau_{\rm{Ross}} = 2/3$ to the radius of the outermost shell of that model, although this outermost shell is determined by an arbitrary choice of the minimum $\tau_{\rm{Ross}}$ when the model is computed.  The quality of fit to the interferometric data is sensitive to the Rossland angular diameter, as was noted by \cite{Wittkowski2004}.  The minimum $\chi^2$ values are shown in Fig. \ref{f1} as a function of limb--darkened and Rossland angular diameter for each model for the case of $\psi$ Phe. The best--fit Rossland angular diameter is well constrained by interferometry and the values of the Rossland and limb--darkened angular diameters, along with the uncertainty of the fit, for each star is given in Table \ref{t1}.  The Rossland angular diameter has a smaller uncertainty than the limb--darkened angular diameter for each star because the fits to the interferometric data produce less variation of the Rossland angular diameter than the limb--darkened angular diameter.  This variation is clearly shown in Fig. \ref{f1}.

   \begin{figure}
   \centering
   \includegraphics[]{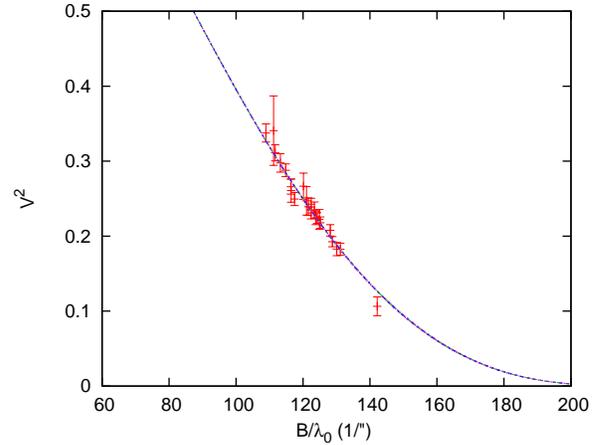}
      \caption{The visibilities calculated for three \textsc{SAtlas} models and the interferometric data for  $\gamma$ Sge. The dotted line refers to the model stellar atmosphere for $L = 700L_\odot$, M = $1.9M_\odot$, and $R = 30R_\odot$, the dot--dashed line represents a model with $L = 550L_\odot$, M = $1.4M_\odot$, and $R = 50R_\odot$, and the dashed line is for $L = 400L_\odot$, M = $1.0M_\odot$, and $R = 70R_\odot$. 
              }
         \label{f3}
   \end{figure}

\begin{figure}
   \centering
   \includegraphics[]{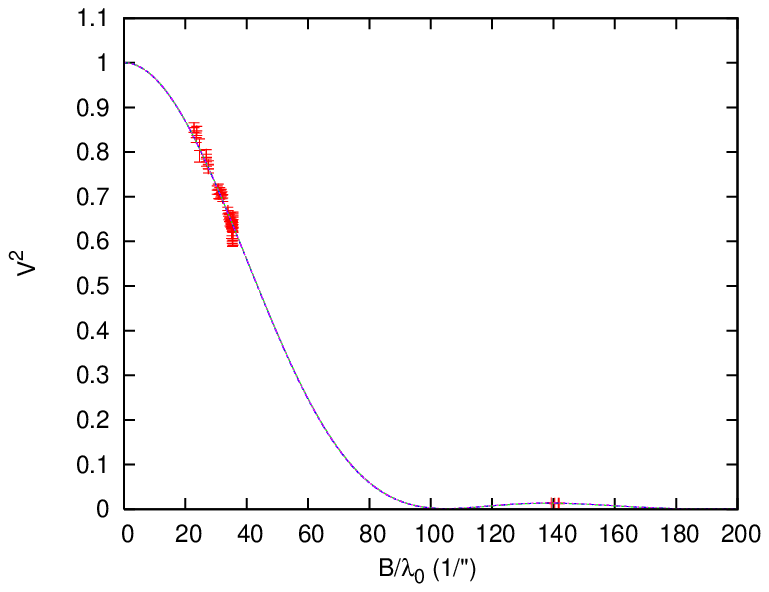} 
   \includegraphics[]{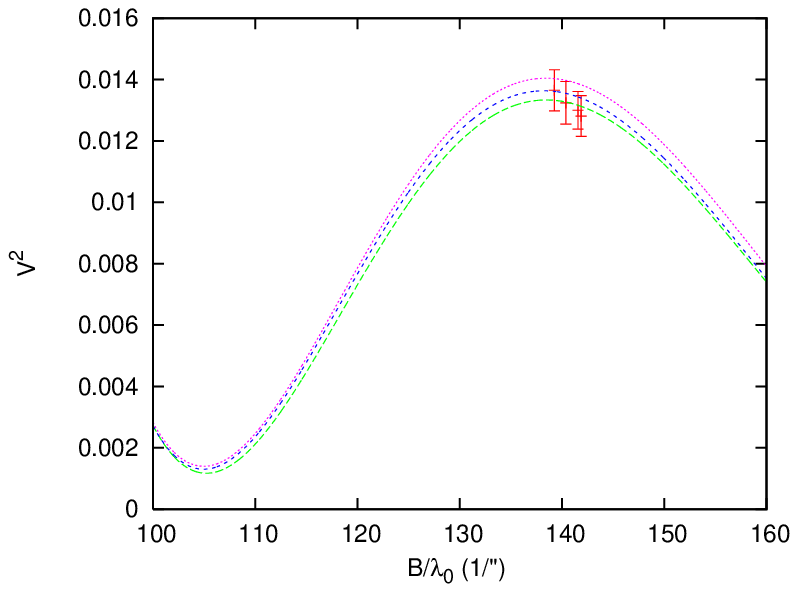}
      \caption{(Top) The visibilities calculated for three \textsc{SAtlas} models and the interferometric data for $\alpha$ Cet. (Bottom) Close--up of the second lobe of the visibility amplitude.  The dotted line refers to the model stellar atmosphere for $L = 2000L_\odot$, M = $3.0M_\odot$, and $R = 60R_\odot$, the dot--dashed line represents a model with $L =1500L_\odot$, M = $2.3M_\odot$, and $R = 80R_\odot$, and the dashed line is for $L = 1000L_\odot$, M = $1.5M_\odot$, and $R = 100R_\odot$. 
              }
         \label{f4}
   \end{figure}

The model visibilities are shown in Fig. \ref{f2} for $\psi$ Phe, with a close--up of the second lobe, Fig. \ref{f3} for $\gamma$ Sge, and Fig. \ref{f4} for $\alpha$ Ceti with a close--up of the second lobe shown.  The  displayed model visibilities are chosen to be the smallest, middle and largest luminosity and gravity from each cube of models.  The model visibilities for $\gamma$ Sge and $\alpha$ Cet agree with the results of \cite{Wittkowski2006b, Wittkowski2006a}. For $\gamma$ Sge, there is not enough information to test the limb--darkening of the model atmospheres.   While the model visibilities using \textsc{SAtlas} and \textsc{Phoenix} both fit the observed interferometric data well within the uncertainties of the observations, the minimum value of the $\chi^2$ from fitting the \textsc{SAtlas} models is smaller than that using the \textsc{Phoenix} models.  This implies that there are small differences in the model atmosphere intensity structures predicted from each program.  It is not obvious why the predictions using \textsc{Phoenix} models and \textsc{SAtlas} models differ in this case.

\begin{figure}[t]
   \centering
 \includegraphics[]{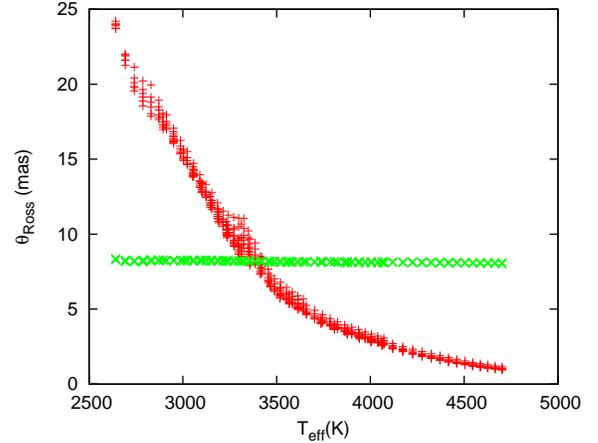}
      \caption{The best--fit effective temperature as a function of Rossland angular diameter found by fitting spectrophotometry (+'s) and interferometry (x's) for $\psi$ Phe.
              }
         \label{f5}
   \end{figure}

Having fit the interferometric data, we next fit the broadband spectrophotometric data. One concern about fitting stellar atmosphere models to spectrophotometry is that the effective temperature and Rossland angular diameter are related.  In Fig. \ref{f5}, we compare the best--fit values of Rossland angular diameter as a function of the best--fit effective temperature for each model for fitting $\psi$ Phe. The predicted angular diameter is almost constant with respect to effective temperature for the fit to interferometry. Therefore the interferometric fit of the Rossland angular diameter is used to constrain the predicted effective temperature.  This process is repeated for the other two stars and the best--fit effective temperatures are given in Table \ref{t1}.

\begin{figure}
   \centering
 \includegraphics[]{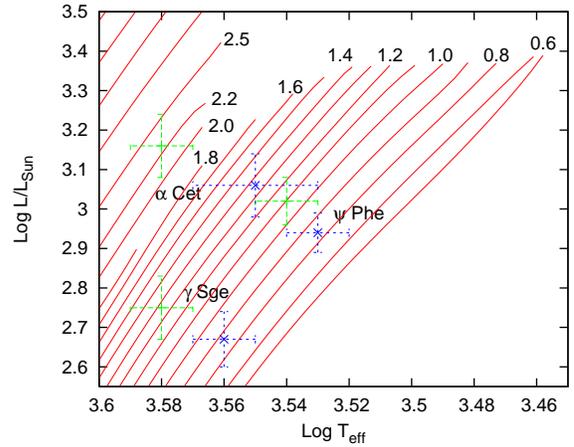}
      \caption{Comparison of the derived effective temperatures and luminosities of the three stars relative the derived parameters from previous analyses. The lines are evolutionary tracks from \cite{Girardi2000}, the points given by x's are the predicted effective temperatures and luminosities found using the \textsc{SAtlas} models and the plusses are the values found using \textsc{Phoenix} models.
              }
         \label{f6}
   \end{figure}

The next step is to calculate the linear radius and luminosity of the stars.  Using the HIPPARCOS parallax with the predicted Rossland angular diameter, we determine the radius of each star, the values of the parallax and radius are given in Table \ref{t1}.  The stellar luminosity is determined using the effective temperature and radius, $L = 4\pi R^2 \sigma T_{\rm{eff}}^4$.  The best--fit effective temperature and luminosity for each star is plotted in Fig. \ref{f6} along with evolutionary tracks from \cite{Girardi2000} to determine the mass and gravity of each star. 

\section{Conclusions}
The purpose of this work is to test the spherical version of the \textsc{Atlas} program. By fitting interferometric and spectrophotometric observations using model stellar atmospheres, we derive the properties of $\psi$ Phe, $\gamma$ Sge, and $\alpha$ Cet.  The derived parameters are compared to earlier results.  The  angular diameters that are determined by fitting model stellar atmospheres to interferometric observations are the same, except for $\alpha$ Cet, for which we predict an angular diameter that is $0.1$ $\rm{mas}$ smaller. The differences in predicted angular diameter are likely due to differences in limb--darkening of the model atmospheres generated using \textsc{SAtlas} and \textsc{Phoenix}.  Also our minimum $\chi^2$ value of the fit of the limb--darkened angular diameter for $\psi$ Phe, $1.66$, is smaller than the predicted value of $\chi^2$ from the spherically symmetric \textsc{Phoenix} models, $1.8$.  The \textsc{SAtlas} models for $\psi$ Phe that best fit the \emph{interferometric} data have effective temperatures in the range of $4000$ to $4500$ $\rm{K}$, while the \textsc{Phoenix} models from \cite{Wittkowski2004} are all $3550$ and $3600$ $\rm{K}$.  The fit of the \textsc{SAtlas} models to \emph{spectrophotometric} observations predict an effective temperature of $3415$ $\rm{K}$ conflicting with the prediction of a higher temperature. For the effective temperature range of $3400$ to $3800$ $\rm{K}$, the minimum $\chi^2$ values predicted by fitting \textsc{SAtlas} models are $1.70$ to $1.72$, similar to the values found using plane--parallel \textsc{Atlas} models.

The fit of the \textsc{SAtlas} models to broadband spectrophotometric observations are used to determine the effective temperatures of the three stars because spectrophotometry is much more sensitive to the effective temperature than interferometry. The predicted effective temperatures are smaller than those found in the previous works, but this difference is due to the methods used for determining the effective temperature, \emph{not} differences in the stellar atmosphere programs. If we use the method from \cite{Wittkowski2004} where the bolometric flux $f_{\rm{bol}}$ is calculated by integrating the spectrophotometric data and then $T_{\rm{eff}}^4 = 4f_{\rm{bol}}/(\sigma \theta_{\rm{Ross}}^2)$, we would predict similar effective temperatures.  Any variation in that case would be due to differences in the calculation of the bolometric flux and differences in the angular diameter.  This suggests the effective temperatures here would differ by at most $1\%$.   Our results also differ because of the spectrophotometric data used.  The  difference is smallest for $\psi$ Phe because the \textsc{Phoenix} and \textsc{SAtlas} fits both use the same spectrophotometric data for the fitting and agree within the uncertainty.  For $\gamma$ Sge, \cite{Wittkowski2006a} complement the \cite{Johnson1975} data with narrow--band data from \cite{Alekseeva1997} while for $\alpha$ Cet, \cite{Wittkowski2006b} use optical \citep{Glushneva1998b, Glushneva1998a} and infrared \citep{Cohen1996} spectrophotometry.  The effective temperature difference for the remaining two stars are about $150$ to $200$ $\rm{K}$. 

Because the angular diameters are similar, the predicted radii are also similar for both fits with \textsc{SAtlas} and \textsc{Phoenix}.  Therefore the differences between predicted luminosities are due to differences in effective temperatures and thus due to the different methods for determining the effective temperatures.  The lower effective temperatures and luminosities imply lower masses when compared to \cite{Girardi2000} evolutionary tracks and smaller gravities.  

The \textsc{SAtlas} models are consistent with previous results. Fitting the models to interferometric and spectrophotometric observations have provided a robust test of the spherically symmetric version of the \textsc{Atlas} program and have shown that the \textsc{SAtlas} program is a powerful tool for studies of stellar atmospheres.

\bibliographystyle{aa}
\bibliography{io}

\end{document}